\documentclass[prb,aps,twocolumn,showpacs,superscriptaddress]{revtex4-1}
\usepackage{hyperref}
\hypersetup{colorlinks=true, citecolor=blue, filecolor= black, linkcolor= blue, urlcolor= blue}
\usepackage{graphicx}
\usepackage{dcolumn}
\usepackage{bm}
\usepackage{float}
\begin{document}


\title{Magnetic borophenes from evolutionary search}

\author{Meng-Hong Zhu}
\email{These coauthors contribute equally to this work.}
\affiliation{Key Laboratory of Weak-Light Nonlinear Photonics and School of Physics, Nankai University, Tianjin 300071, China}

\author{Xiao-Ji Weng}
\email{These coauthors contribute equally to this work.}
\affiliation{Key Laboratory of Weak-Light Nonlinear Photonics and School of Physics, Nankai University, Tianjin 300071, China}

\author{Guoying Gao}
\affiliation{State Key Laboratory of Metastable Materials Science and Technology, Yanshan University, Qinhuangdao 066004, China}

\author{Shuai Dong}
\affiliation{School of Physics, Southeast University, Nanjing 211189, China}

\author{Ling-Fang Lin}
\affiliation{School of Physics, Southeast University, Nanjing 211189, China}

\author{Wei-Hua Wang}
\affiliation{Department of Electronic Science and Engineering, Tianjin Key Laboratory of Photo-Electronic Thin Film Device and Technology, Nankai University, Tianjin 300071, China}

\author{Qiang Zhu}
\affiliation{Department of Physics and Astronomy, High Pressure Science and Engineering Center, University of Nevada, Las Vegas, Nevada 89154, USA}

\author{Artem R. Oganov}
\affiliation{Skolkovo Institute of Science and Technology, 3 Nobel Street, Moscow 143026, Russia}
\affiliation{Moscow Institute of Physics and Technology, Dolgoprudny, Moscow Region 141700, Russia}
\affiliation{International Center for Materials Discovery, Northwestern Polytechnical University, Xi¡¯an 710072, China}

\author{Xiao Dong}
\email{xiao.dong@nankai.edu.cn}
\affiliation{Key Laboratory of Weak-Light Nonlinear Photonics and School of Physics, Nankai University, Tianjin 300071, China}

\author{Yongjun Tian}
\affiliation{State Key Laboratory of Metastable Materials Science and Technology, Yanshan University, Qinhuangdao 066004, China}

\author{Xiang-Feng Zhou}
\email{xfzhou@nankai.edu.cn}
\email{zxf888@163.com}
\affiliation{Key Laboratory of Weak-Light Nonlinear Photonics and School of Physics, Nankai University, Tianjin 300071, China}
\affiliation{State Key Laboratory of Metastable Materials Science and Technology, Yanshan University, Qinhuangdao 066004, China}

\author{Hui-Tian Wang}
\affiliation{Key Laboratory of Weak-Light Nonlinear Photonics and School of Physics, Nankai University, Tianjin 300071, China}
\affiliation{National Laboratory of Solid State Microstructures and Collaborative Innovation Center of Advanced Microstructures, Nanjing University, Nanjing 210093, China}

\begin{abstract}
\noindent A computational methodology based on \textit{ab initio} evolutionary algorithms and the spin-polarized density functional theory was developed to predict two-dimensional (2D) magnetic materials. Its application to a model system borophene reveals an unexpected rich magnetism and polymorphism. A stable borophene with nonzero thickness is an antiferromagnetic (AFM) semiconductor from first-principles calculations, which can be further tuned into a half metal by finite electron doping. In this borophene, the buckling and coupling among three atomic layers are not only responsible for magnetism, but also result in an out-of-plane negative Poisson's ratios under uniaxial tension, making it the first elemental material possessing auxetic and magnetic properties simultaneously.
\end{abstract}



\maketitle
\section{INTRODUCTION}
Two-dimensional (2D) magnetic materials have attracted huge interest owing to their potential applications in spintronics and data storage. \cite{R01,R02,R03,R04,R05,R06,R07,R08,R09,R10,R11,R12} The ultimate thinness changes the physical properties dramatically compared to the corresponding bulk materials. Although a number of 2D magnetic materials have been proposed theoretically, \cite{R09,R10,R11,R12,R13,R14} few of them were synthesized experimentally. Recent experimental studies found that 2D forms of CrI$_{3}$ and Cr$_{2}$Ge$_{2}$Te$_{6}$ can inherit the ferromagnetic (FM) orders at low temperatures from their bulk forms. \cite{R01,R02} Strikingly, even for paramagnetic (PM) bulk material, i.e., VSe$_{2}$, its monolayer form shows an unexpected strong room-temperature ferromagnetism on van der Waals substrates. \cite{R03} Stimulated by these exiting experimental reports, 56 new magnetically ordered monolayer structures were predicted from high-throughput computation to be exfoliable from known magnetic bulk materials. \cite{R04} However, discovering new 2D magnetic materials beyond exfoliation from the parent bulk compounds is still governed by trial and error approaches. Among the light element-based 2D materials, graphene is not magnetic, \cite{R15} but may be a promising material for spintronics by optimizing defects, substrate, or adsorbing hydrogen atoms. \cite{R16,R17,R18,R19} Boron, with only three valence electrons, is electron deficient, resulting in the formation of multicenter B-B bonds and rich polymorphism with great chemical complexity in borophenes. \cite{R20,R21,R22,R23,R24,R25,R26,R27,R28,R29,R30,R31} Various borophenes have been successfully synthesized on Ag, Cu, or Al substrates under ultrahigh-vacuum conditions. \cite{R32,R33,R34,R35,R36} The atomically thin borophenes display remarkable properties, such as the emergence of superconductivity, massless Dirac fermions, superior transport and mechanical properties. \cite{R20,R21,R22,R23,R24,R25,R26,R27,R28,R29,R30,R31,R32,R33,R34,R35,R36} However, the phase diagram of borophene remains largely unexplored due to lack of efficient structure prediction method. With the goal of uncovering more 2D mateials with superior magnetic properties besides magnetic doping that have been experimentally confirmed as one of simple and efficient methods, \cite{R29,R30,R31} we developed a new strategy based on evolutionary structure prediction and explored the magnetic borophenes in a systematic manner.

\begin{figure}[t]
\begin{center}
\includegraphics[width=8cm]{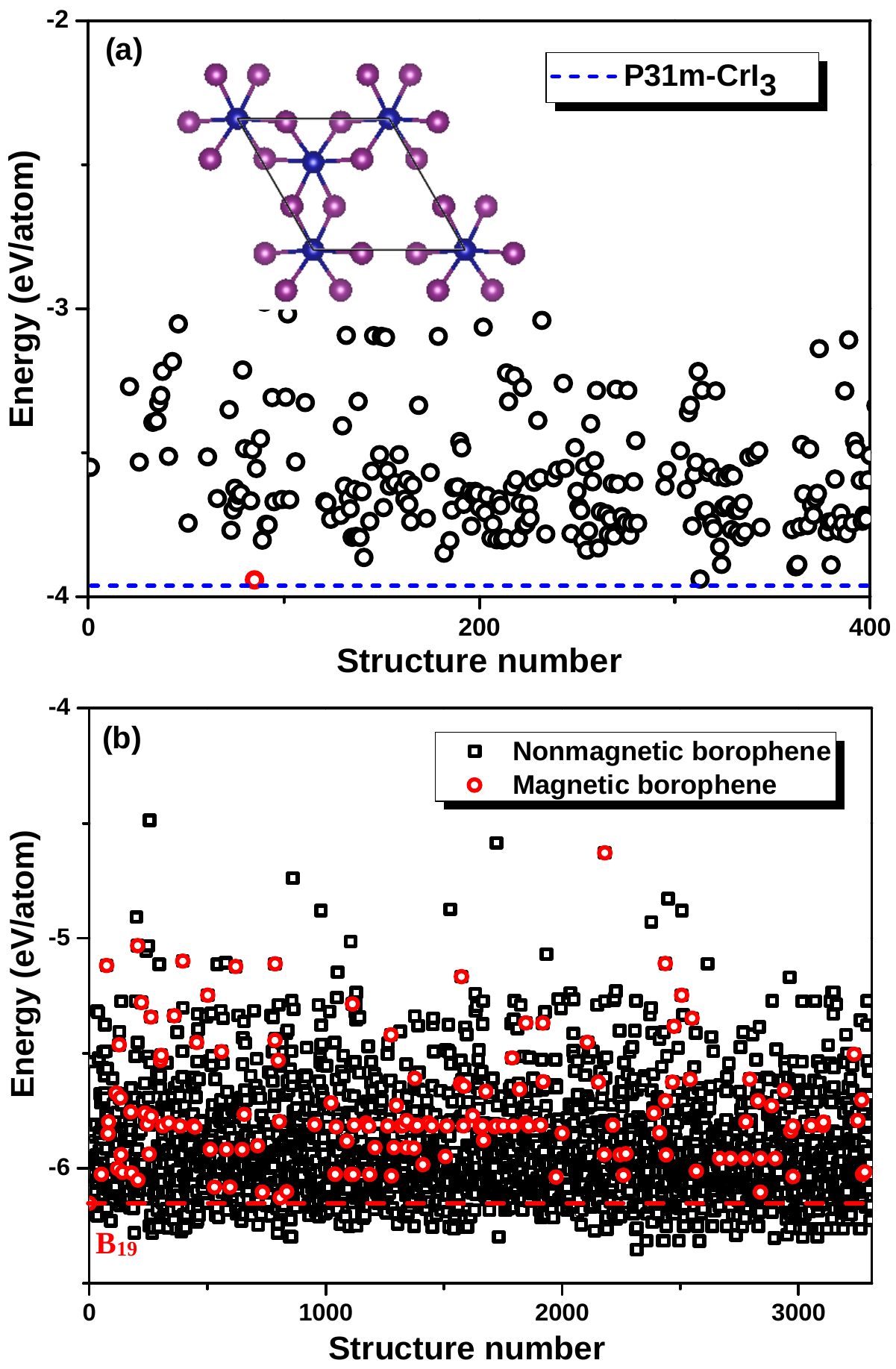}
\caption{%
(color online) (a) Energy evolution for 2D CrI$_{3}$ system during the magnetic evolutionary searching. All of structures are magnetic and the inset shows the most stable structure of $P3m1$-CrI$_{3}$. (b) Energy distribution of borophenes from the magnetic evolutionary searching.}
\end{center}
\end{figure}

\section{METHOD}
To search for the stable 2D magnetic structures, we further developed a new computational scheme based on the \textit{ab initio} evolutionary algorithm \textsc{uspex} \cite{R37,R38,R39,R40} combined with the spin-polarized density functional theory (DFT). The initial structures are produced with a randomly assigned layer group symmetry and a user-defined initial thickness of 2~{\AA}, while the initial magnetic moment was generated by random numbers, which are classified as 0 (0,0.1], 1 (0.1, 0.3], 4 (0.3, 0.5], $\pm 1$ (0.5, 0.75], and $\pm 4$ (0.75, 1.0], and then the structures are treated as nonmagnetic (NM) order, FM with low spin (FM-LS), FM with high spin (FM-HS), AFM with low spin (AFM-LS), and AFM with high spin (AFM-HS), \cite{R41} respectively. The structures with odd number of atoms are supposed to have FM order. The ratio of structures with different magnetic orders (NM, FM-LS, FM-HS, AFM-LS, and AFM-HS) can be either uniform or user-defined. The relaxed energies associated with the magnetic moment were used as a criterion for parent structures selection to generate new structures by various evolutionary operators, such as random structure generator, heredity and mutations (including lattice, coordination, and spin heredity/mutations). For each relaxed structure, we sum over the (absolute) values of the magnetic moments for all atoms ($\sum M_{i}$ and $\sum |M_{i}|$). If $\sum |M_{i}|$ is close to zero ($< 0.03 \times N_{atoms}$, where $N_{atoms}$ is the number of magnetic atoms), we assign it to be NM; otherwise, the structure would be either AFM ($|\sum M_{i}| < 0.25 \times N_{atoms}$) or FM ($|\sum M_{i}| \geq 0.25 \times N_{atoms}$). We also assign the structure to be LS \{if $\max(|M_{i}|) < 1.5$\} or HS \{$\min(|M_{i}|) > 1.5$\}. Otherwise, the structure would be labelled as a hybrid HSLS. We limited our structural search within an unit cell of 5-20 atoms with a vacuum of 20~{\AA}. Structure relaxations and total energy calculations were performed with the projector-augmented wave \cite{R42} (PAW) method as implemented in the VASP package, \cite{R43} and the exchange-correlation energy was computed within the generalized gradient approximation (GGA) with the functional of Perdew, Burke, and Ernzerhof (PBE). \cite{R44} The plane-wave cutoff energy of 600 eV and uniform $\Gamma$-centered k-points grids with the resolution of $2 \pi \times 0.025$ ~\AA$^{-1}$ were used. The convergence for terminating the electronic self-consistency cycle and the force criterion for structure relaxation were set at 10$^{-6}$ eV and 10$^{-2}$ eV/{\AA}. Phonon dispersion curves were calculated by the supercell method (the $2 \times 2 \times 1$ supercell for NM, FM and AFM states) using the \textsc{phonopy} package \cite{R45} with the energy convergence of 10$^{-7}$ eV.

\begin{figure}[t]
\begin{center}
\includegraphics[width=8cm]{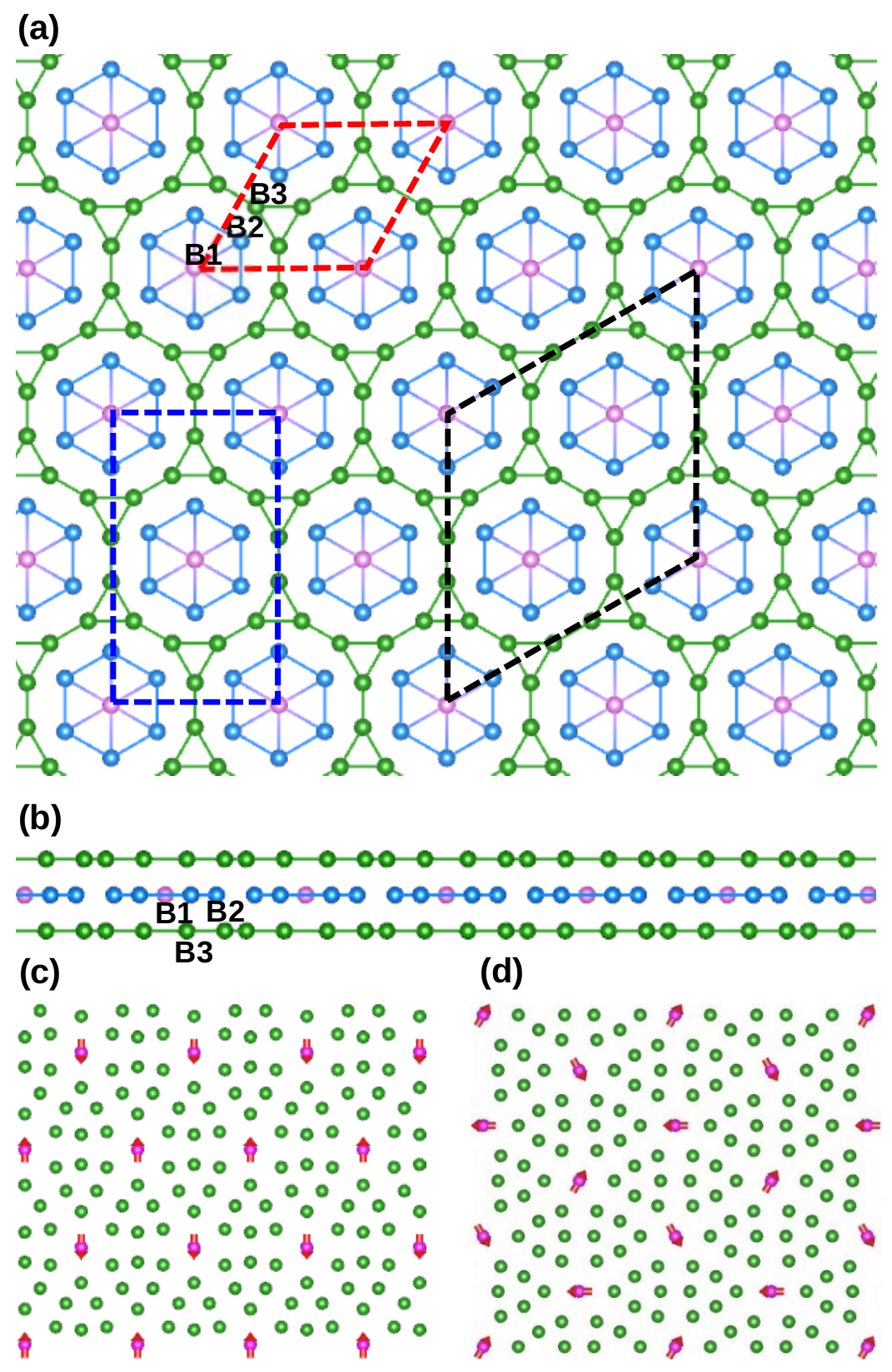}
\caption{%
(color online) (a) and (b) Top and side views of $19-P6/mmm$ borophene. The partial bonds among three layers are removed for clarity. Three inequivalent atomic positions (B1, B2, and B3) of NM structure are labelled. The dotted red, blue and black lines indicate the lattices of NM, striped-AFM, and non-collinear $120^\circ$ AFM (ncl-AFM) structures, respectively. (c) and (d) striped-AFM and ncl-AFM, the pink arrows indicate the relative directions of their magnetic moments.}
\end{center}
\end{figure}

\section{RESULTS AND DISCUSSION}


\begin{figure*}
\begin{center}
\includegraphics[width=1.5\columnwidth]{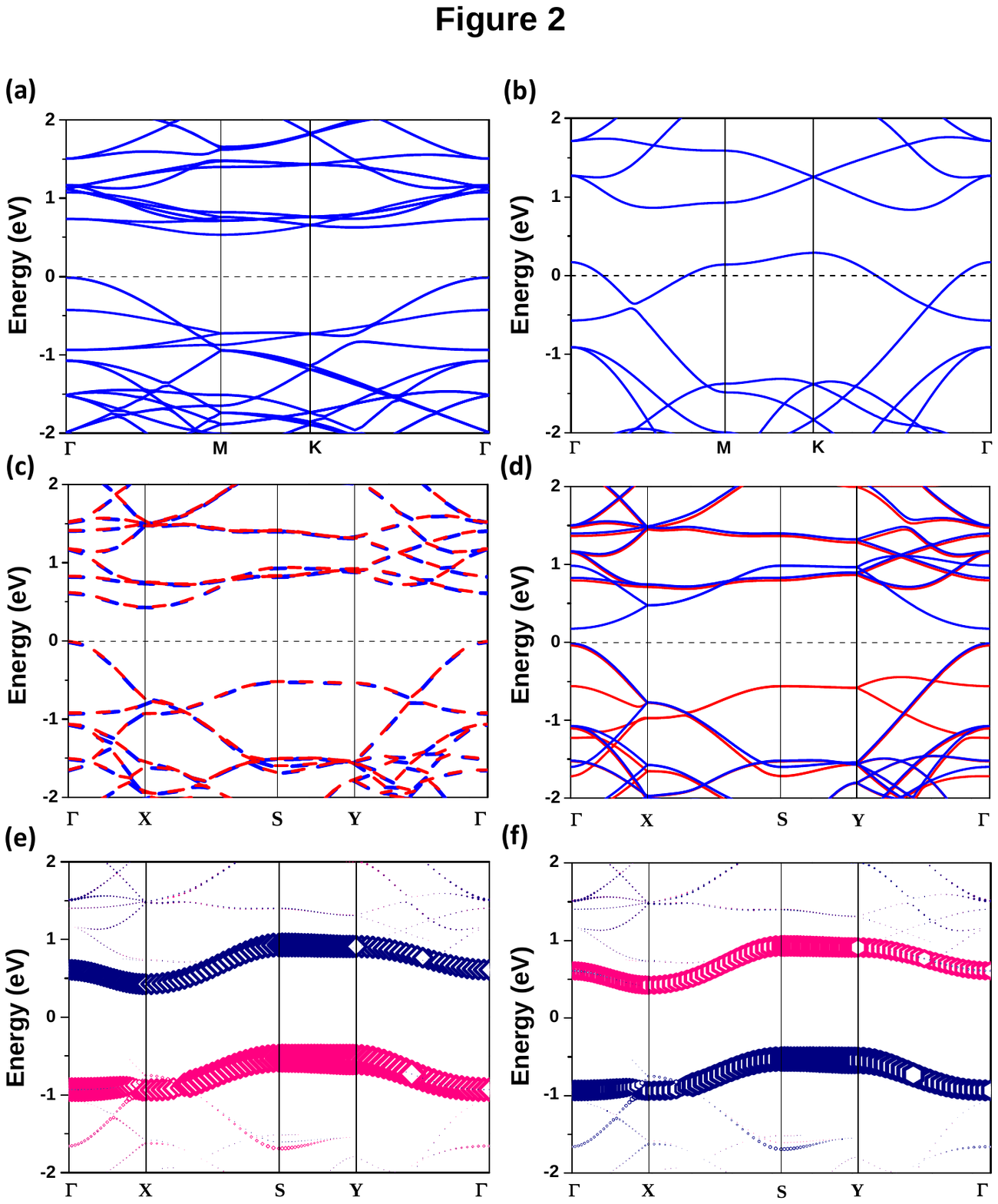}
\end{center}
\caption{\label{fig3}
(color online) Band structures of $19-P6/mmm$ borophene with (a) ncl-AFM, (b) NM, (c) striped-AFM, (d) FM structures. (e) The atomic orbital resolved band structures of striped-AFM $19-P6/mmm$ borophene. Two flat bands dominated by the $p_z$ orbitals primarily originate from one type of B1 atoms. The highest valence band with majority spin is colored in pink, and the lowest conduction band with minority spin is colored in dark blue. (f) These two flat bands mainly originate from the other magnetic B1 atoms of striped-AFM state, but their spin electrons behaves oppositely.}
\end{figure*}

To test the reliability and accuracy of this new method, we first investigated 2D CrI$_{3}$ system. As shown in Fig. 1a, our calculations demonstrated that the most stable structure has ferromagnetic order and $P31m$ symmetry, and its lattice constants and magnetization (3 $\mu_{B}$ per Cr atom) are in good agreement with the experimental results. \cite{R02} Compared with high-throughput computational screening, \cite{R04,R13,R14} this search uncovered many new metastable magnetic structures which do not have analogous known parent bulk materials in the database, suggesting that our search is not biased by the database and thus offers a more complete sampling of the configurational space. We then applied this method to borophene. Both magnetic and nonmagnetic borophenes are unveiled in Fig. 1b. Fewer than 300 magnetic borophenes were predicted among 3500 structures in total, after the removal of duplicate structures which were found more than once in the search. We also note that magnetic calculations may be numerically sensitive to the choice of computational parameters, and magnetism of some planar monolayers may disappear in more converged calculations because high density of states (DOS) at the Fermi level can be lowered by structural (Peierls) distortion rather than spin polarization. Since elemental boron prefers NM state, all magnetic borophenes are metastable phases with respect to its ground-state structure. However, rigorous convergence of calculations for all structures would be prohibitive because it is beyond our computational resource. Hence, we focus on these low-energy structures in which ferromagnetism persists, as they are more likely to be synthesized on a suitably chosen substrate. The most stable magnetic structure contains 19 atoms per unit cell (Fig. 2) and is designated as $19-P6/mmm$ borophenes according to its symmetry. For $19-P6/mmm$ borophene with NM order (NM $19-P6/mmm$ borophene), the lattice parameters are a = b = 6.033 ~{\AA}, and c = 19.97 ~{\AA}. Three inequivalent atomic positions are B1 (0.0,0.0,0.5), B2 (0.362,0.181,0.5), and B3 (0.576,0.153,0.434), which form triatomic 2.64 ~{\AA} thick layers (see Figs. 2a and 2b). The middle layer, through which a mirror plane passes, consists of isolated hexagonal B$_{7}$ clusters (meanwhile, leading to large empty spaces, see Fig. 2a), while the top and bottom layers, are composed of triangular units, and they connect together to form dodecagonal vacancies on top of the B$_{7}$ clusters, which forms particular symmetric voided structure.

\begin{figure}[t]
\begin{center}
\includegraphics[width=8.0cm]{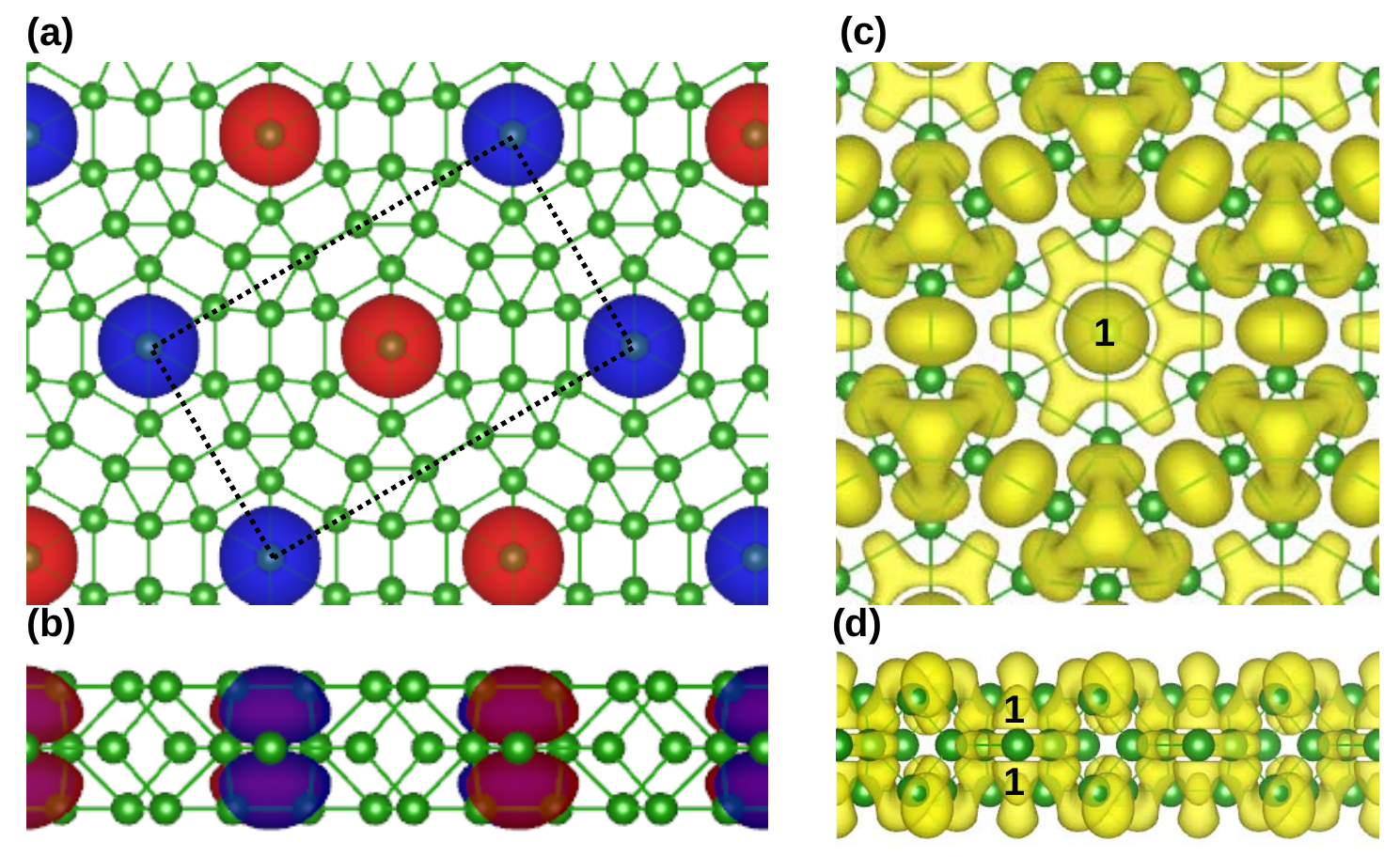}
\caption{%
(color online) (a) and (b) Top and side views of spin charge density of striped-AFM $19-P6/mmm$ borophene. The majority-spin charge density is colored in red, while the minority-spin charge density is colored in blue. The dotted lines indicate the rectangular lattice of striped-AFM structure. (c) and (d) Top and side views of ELF of FM $19-P6/mmm$ borophene. The unpaired electrons are labeled as number 1 in c and d.}
\end{center}
\end{figure}

The calculated total energies for the $19-P6/mmm$ borophene with striped-AFM (Fig. 2c), non-collinear $120^\circ$ AFM order (ncl-AFM, see Figs. 2a and 2d), FM, and NM states are -6.162, -6.162, -6.161, and -6.152 eV/atom, which are higher in energies than previously synthesized 2-Pmmn (-6.19 eV/atom), \cite{R32} $\chi_{3}$ (-6.24 eV/atom), \cite{R34} $\beta_{12}$ (-6.23 eV/atom), \cite{R34} and $\alpha$-sheet structures (-6.28 eV/atom), \cite{R20} but are lower in energy than graphene-like borophene (-5.42 eV/atom), \cite{R35} indicating that $19-P6/mmm$ borophene is a metastable structure with the ground-state AFM order. There is only one magnetic atom per unit cell of FM $19-P6/mmm$ borophene (Fig. 2a), which is located at the center of B$_{7}$ clusters (B1 atoms). The local magnetic moment of each B1 atom is about 1.0 $\mu_{B}$. The geometry of B1 atoms forms a magnetic triangular lattice and thus many configurations include FM, striped-AFM and ncl-AFM orders are considered (Fig. 2). The calculations show that ncl-AFM phase, which are considerably, by 2.2 meV per 19 atoms, and 24.5 meV per 19 atoms, lower in energy than those of striped-AFM and FM orders. Therefore, ncl-AFM borophene is the ground-state structure, which is an indirect semiconductor with bandgap of 0.54 eV (Fig. 3a). Moreover, by mapping the DFT energy difference to the classical spin model, the nearest-neighbor exchange parameter $J_{1}$ is estimated to be -5.45 meV and the next nearest-neighbor exchange parameter $J_{2}$ is $\sim$ -0.14 meV. $J_{1}$ is much bigger than $J_{2}$, indicating that $19-P6/mmm$ borophene has a ground-state ncl-AFM structure, which is dominated by the AFM $J_{1}$ in the triangular lattice. However, under minor electron doping, ncl-AFM phase is less stable than striped-AFM and FM phases, and there is even an AFM-FM transition when doping concentration is higher than $\sim$ 0.006 electrons per atom ($3.6 \times 10^{13}$ $cm^{-2}$). Hence the ncl-AFM structure will not be discussed further in subsequent discussions for the doped cases.

\begin{figure}[t]
\begin{center}
\includegraphics[width=8.5cm]{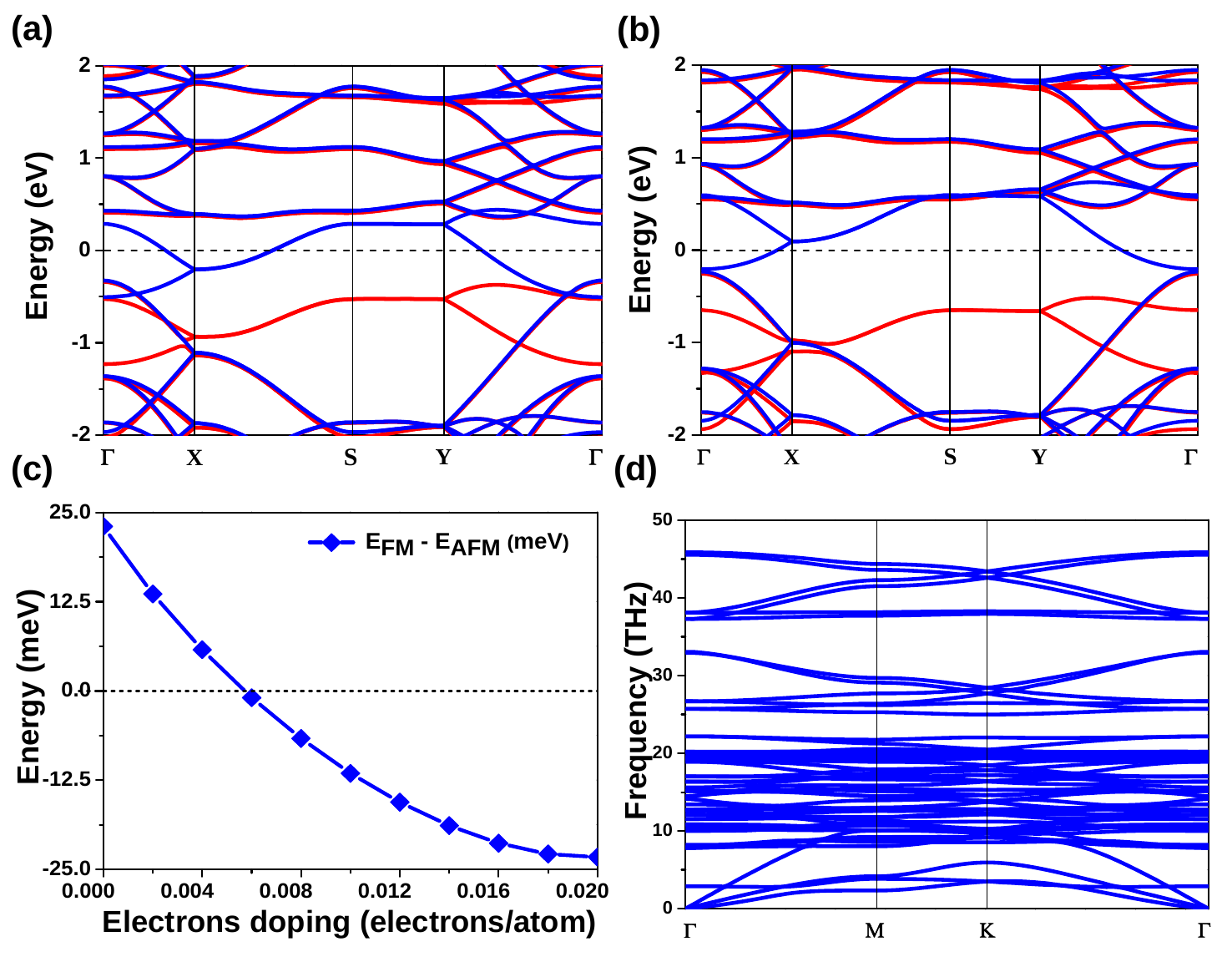}
\caption{%
(color online) (a) Band structure of FM $19-P6/mmm$ borophene with the electron doping concentration of $1.2 \times 10^{14}$ $cm^{-2}$. (b) Band structure with the electron doping concentration of $4.8 \times 10^{13}$ $cm^{-2}$. (c) The variation of relative energy difference between FM and striped-AFM phases by finite electron doping. (d) Phonon dispersion curve of FM $19-P6/mmm$ borophene with the doping concentration of 0.02 electrons per atom.}
\end{center}
\end{figure}


Electronic band structure calculations show that NM $19-P6/mmm$ borophene is metallic (Fig. 3b). In contrast, striped-AFM and FM $19-P6/mmm$ borophenes are semiconductors with an indirect bandgap of 0.41 eV and a direct bandgap of 0.18 eV, respectively (Figs. 3c and 3d). The electronic stability of different states (NM metal $<$ FM or AFM semiconductor) is consistent with the energetic stability. To explore the physical origin of magnetism in striped-AFM borophene, we analyzed the atomic orbital-resolved band structures as shown in Figs. 3e and 3f. In Fig. 3e, two flat bands around the Fermi level are primarily contributed by the $p_z$ orbitals of magnetic B1 atoms. Between them, one flat band with majority spin is colored in pink, while the other with minority spin is colored in dark blue. Interestingly, there also exist two more flat bands in Fig. 3f, which mainly originate from the other B1 atoms, but their spin characters are opposite. That is, two highest valence bands or two lowest conduction bands (flat bands) with opposite spin electrons are offseted each other (Figs. 3e and 3f) and thus, the whole structure exhibits AFM order. As mentioned above, these flat bands are dominated by the unpaired electrons ($p_z$ orbitals), which can be confirmed by the spin charge density distribution (Figs. 4a and 4b). Large bubble-like spin density maxima are localized on top of B1 atoms (center of B$_{7}$ clusters). These bubbles represent the majority-spin (colored in red) and minority-spin electrons (colored in blue) and have the same size and shape due to mirror symmetry. The special arrangement of spin densities among B1 atoms in the plane are responsible for striped-AFM order. In addition, the electron localization function (ELF) of FM $19-P6/mmm$ borophene in Figs. 4c and 4d shows that each B1 atom has an unpaired electron, and consequently has a local magnetic moment of 1 $\mu_{B}$.

\begin{figure}[t]
\begin{center}
\includegraphics[width=8.0cm]{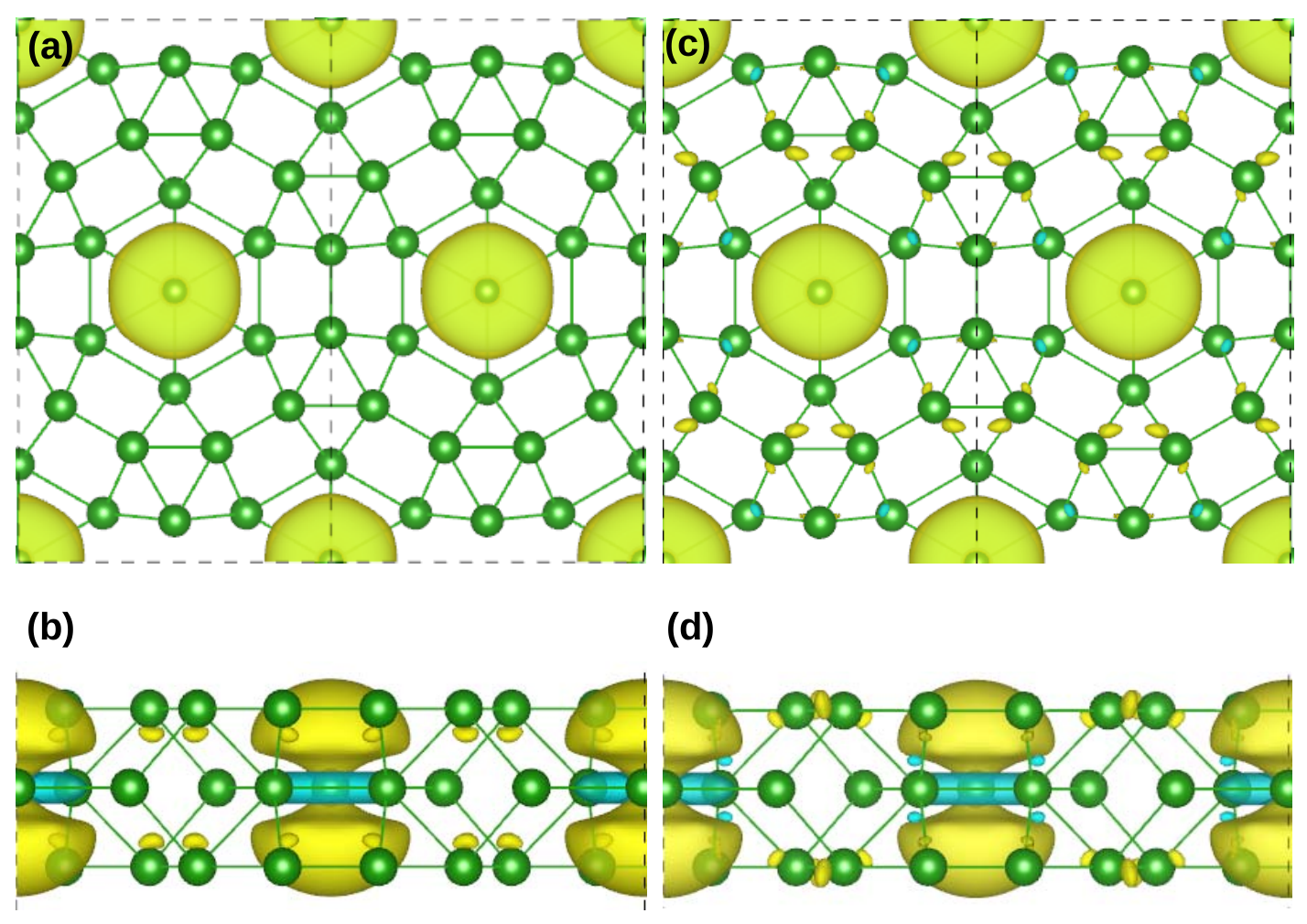}
\caption{%
(color online) Top and side views of charge density difference in $19-P6/mmm$ borophene between the doped (0.02 electrons per atom) and undoped cases. (a) and (b) FM, (c) and (d) striped-AFM states.}
\end{center}
\end{figure}

\begin{figure}[t]
\begin{center}
\includegraphics[width=8cm]{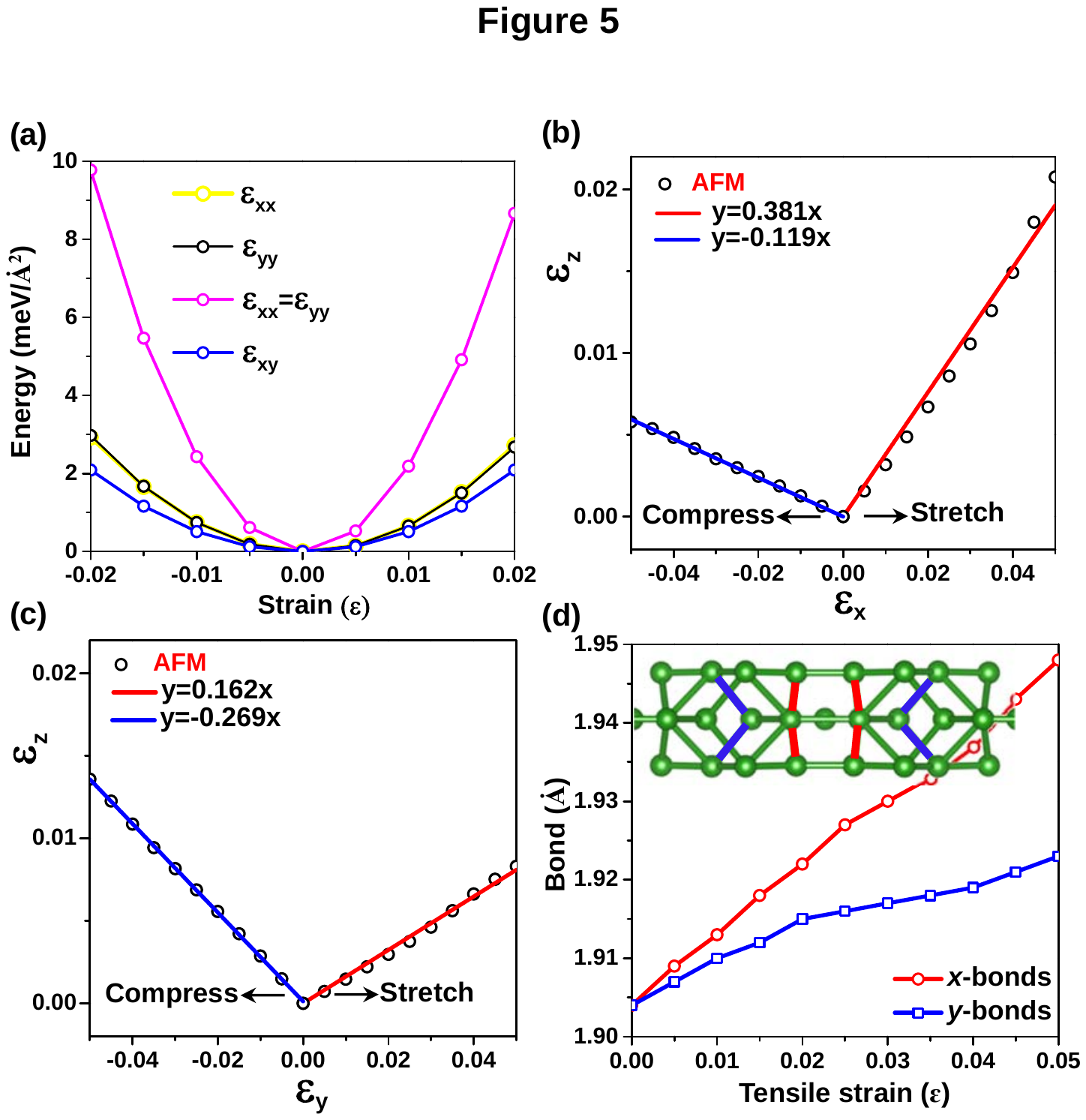}
\caption{%
(color online) Striped-AFM $19-P6/mmm$ borophene under strain. (a) Calculated total energy vs. strain relation. (b) The Poisson's ratios as a function of uniaxial deformation of striped-AFM structure along $x$ direction. (c) Poisson's ratios as a function of uniaxial deformation of AFM structure along $y$ direction. (d) The lengths of specific B2-B3 bonds as a function of uniaxial tensile strain. The inset shows the $x$-bonds and $y$-bonds, which are colored in red and blue respectively.}
\end{center}
\end{figure}

FM $19-P6/mmm$ borophene is a direct narrow-gap semiconductor. As see in Fig. 3d, the lowest conduction band with minority spin is very close to the Fermi level along $\Gamma$--X and $\Gamma$--Y directions, while the conduction band with majority spin is relatively far from the Fermi level. Obviously, it may be tuned from the FM semiconductor to a FM half metal by shifting the Fermi level upward a little, which can be realized by electron doping. Note that a charge doping of $10^{15}$ $cm^{-2}$ had already been experimentally achieved in some 2D materials by a gate-voltage. \cite{R46} For the electron doping concentration of 0.02 ($1.2 \times 10^{14}$ $cm^{-2}$) and 0.008 ($4.8 \times 10^{13}$ $cm^{-2}$) electrons per atom (Figs. 5a and 5b), the bands in the majority and minority spin channels are separated, suggesting intrinsic ferromagnetism. Most importantly, minority spin channels are completely polarized, indicating that $19-P6/mmm$ borophene is a half-metal whose majority spin electrons behave like semiconductor, while minority spin electrons display metallic conduction. All of these suggest that this structure may have significant potential applications in spintronics. Figure 5c shows the variation of relative energy difference between FM and striped-AFM phases via the electron doping concentration from 0 to $1.2 \times 10^{14}$ $cm^{-2}$. Actually, the $19-P6/mmm$ borophene is a ground-state FM half metal above the concentration of 0.006 electrons per atom. We also calculated the phonon dispersion curves with finite electron doping. The absence of imaginary frequencies in the phonon spectrum (Fig. 5d) is suggesting that the structure remains dynamically stable upon doping. Note that the rectangular $19-P6/mmm$ borophene has a large unit cell and the excess electrons are dominantly distributed on the magnetic B1 atoms in both FM and striped-AFM states (Fig. 6), the weak spurious electrostatic interaction can be cancelled out, so the energy differences between the charged FM and striped-AFM states are correct. \cite{R47}

\begin{figure}[t]
\begin{center}
\includegraphics[width=7.5cm]{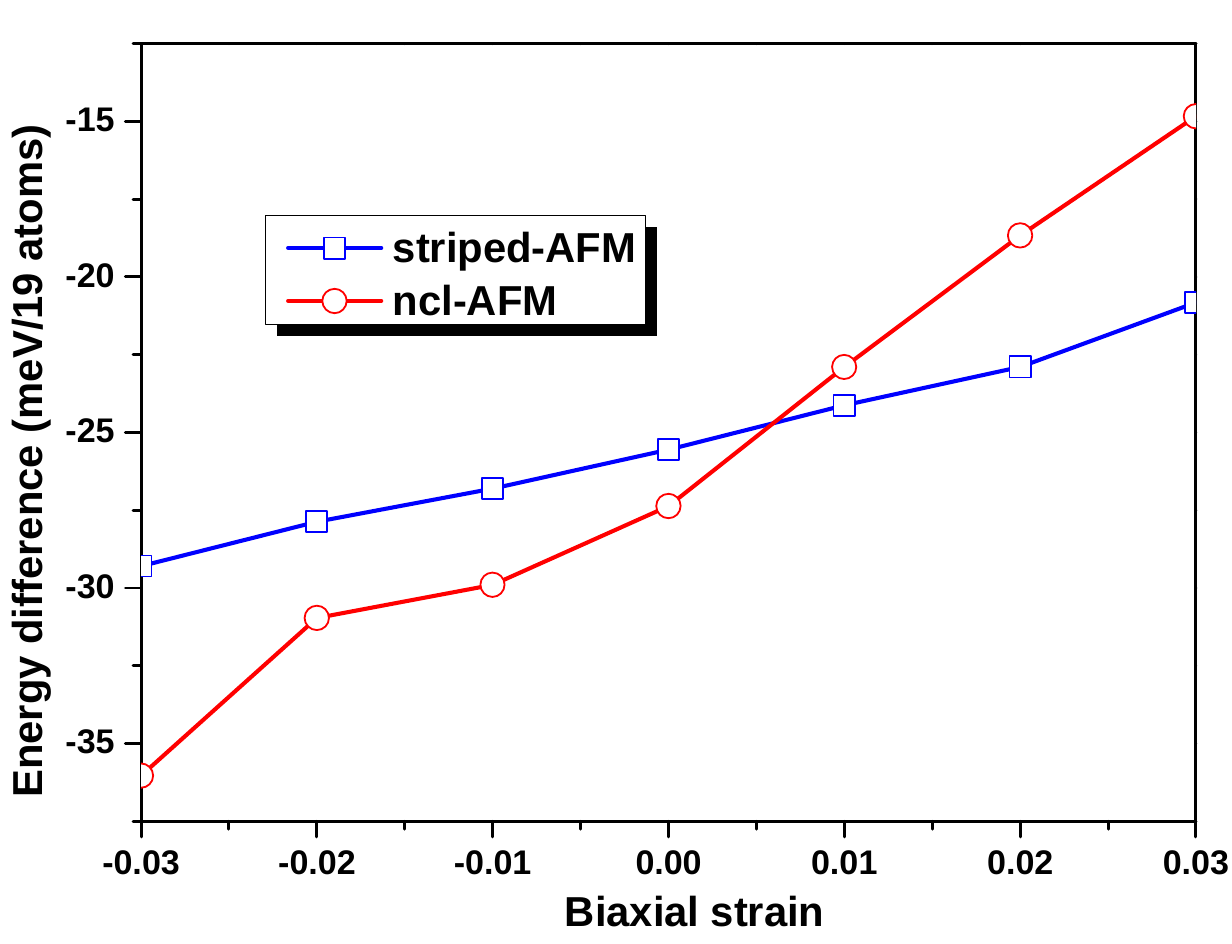}
\caption{%
The energy difference of striped-AFM, ncl-AFM refer to FM phases via the biaxial tensile (negative) or compressive strain (positive).}
\end{center}
\end{figure}

Because of the strong covalent B-B networks, borophenes are usually expected to be hard 2D materials, e.g., the in-plane Young¡¯s modulus of $2-Pmmn$ borophene (398 GPa$\cdot$nm along the a axis), potentially rivals graphene (the hardest 2D material) at 340 GPa$\cdot$nm. \cite{R32,R48} It is natural to study the mechanical properties of $19-P6/mmm$ borophene with different magnetic states. With its reconstructed rectangular lattice, using the standard Voigt notation, the elastic strain energy per unit area can be expressed as\cite{R49} U$_{\epsilon}$ = (1/2)C$_{11}$$\epsilon^2_{xx}$ + (1/2)C$_{22}$$\epsilon^2_{yy}$ + C$_{12}$$\epsilon_{xx}$$\epsilon_{yy}$ + 2C$_{66}$$\epsilon^2_{xy}$, where C$_{11}$, C$_{22}$, C$_{12}$, and C$_{66}$ are the elastic constants, corresponding to second partial derivatives of the energy with respect to strain. The in-plane Young's modulus and Poisson's ratio can be derived from the elastic constants as: E$_{x}$ = (C$_{11}$C$_{22}$--C$_{12}$C$_{21}$)/C$_{22}$, E$_{y}$ = (C$_{11}$C$_{22}$--C$_{12}$C$_{21}$)/C$_{11}$, $\nu_{xy}$ = C$_{21}$/C$_{22}$, and $\nu_{yx}$ = C$_{12}$/C$_{11}$. The calculated C$_{11}$, C$_{22}$, C$_{12}$, and C$_{66}$ for striped-AFM $19-P6/mmm$ borophene are 227, 227, 143, and 42 GPa$\cdot$nm, so the Young¡¯s modulus is equal to $E_{x}$ = $E_{y}$ = 137 GPa$\cdot$nm, and the corresponding Poisson's ratio is equal to 0.63 (Fig. 7a), which are much softer than previously reported structures due to its large empty spaces among three layers. \cite{R32,R50} In sharp contrast to the positive in-plane Poisson's ratio ($\nu_{xy}$ = $\nu_{yx}$ = 0.63), $19-P6/mmm$ borophene, regardless of its magnetic order, possesses an unexpected out-of-plane negative Poisson's ratios (NPR) when tensile strain was applied in the $x$ (parallel to the $a$ axis) and $y$ (parallel to $b$ axis) directions. As shown in Figs. 7b and 7c, the out-of-plane strains $\epsilon_{z}$ of striped-AFM $19-P6/mmm$ borophene have the linear relationship with $\epsilon_{x}$ and $\epsilon_{y}$, respectively. When it was compressed, the data were fitted to the function of $y$ = --0.119(--0.269)$x$ in the $x$ ($y$) direction. Since the out-of-plane Poisson's ratio \cite{R51} is defined as $\nu_{zx}$ = --$\partial$ $\epsilon_{z}$/$\partial$ $\epsilon_{x}$ or $\nu_{zy}$ = --$\partial$ $\epsilon_{z}$/$\partial$ $\epsilon_{y}$, these results represent positive Poisson's ratios in both directions. While for tension, the data were fitted to the function of $y$ = 0.381(0.162)$x$, so they are $\nu_{zx}$ = --0.381 and $\nu_{zy}$ = --0.162, showing NPR effect on the contrary, which are different from those of phosphorene or borophane (NPR for both tension and compression only either in the $y$ or in the $x$ direction). \cite{R52,R53} The length of specific B2-B3 bonds (named as $x$-bonds, colored in red, see Fig. 7d) increases under tensile strain along the $x$ direction, so does for the other specific B2-B3 bonds (named as $y$-bonds, colored in blue, see Fig. 7d) along the $y$ direction. The slope of $x$-bonds is more than that of $y$-bonds, implying that $x$-bonds elongate faster than $y$-bonds upon stretching and resulting in$|$$\nu_{zx}$$|$ $>$ $|$$\nu_{zy}$$|$. Owing to very similar lattice constants among AFM, FM, and NM states, the $\nu_{zx}$ and $\nu_{zy}$ of NM (FM) $19-P6/mmm$ borophene are --0.416 (--0.379) and --0.194 (--0.167) by using the same methods. Therefore, such exotic buckling bonding configurations are responsible for the NPR effect (see the inset of Fig. 7d). At last, the energy difference of striped-AFM, ncl-AFM refer to the FM phases via the biaxial tensile (negative strain) or compressive strain (positive strain) was plotted in Fig. 8. It showed that the striped-AFM structure is the most stable phase among different states between the range of 1\% and 3\%, while the ncl-AFM structure is the most stable one under biaxial tensile strain.

\section{CONCLUSION}
In conclusion, we developed and performed a systematic evolutionary search for stable magnetic borophenes and identified that $19-P6/mmm$ borophene is a stable striped-AFM semiconductor, which not only can be further tuned into a half metal, but also has intrinsic out-of-plane negative Poisson's ratios. The $19-P6/mmm$ borophene may be grown on Au (111) substrate because of the small lattice mismatch ($<$ 5\%), or be made on suitable substrates of transition-metal borides. \cite{R54} Our calculations show that the intrinsic magnetism is destroyed by the interaction between $19-P6/mmm$ borophene and the substrate, i.e, Au (111). However, if the freestanding form can be realized from experiments, the outstanding properties make it a promising candidate in spintronics and nanoelectromechanical devices simultaneously.

\textbf{ACKNOWLEDGMENTS} This work was supported by the National Science Foundation of China (Grants 11674176 and 11874224), the Tianjin Science Foundation for Distinguished Young Scholars (Grant No. 17JCJQJC44400). A.R.O. thanks Russian Science Foundation (Grant No. 16-13-10459), and the Foreign Talents Introduction and Academic Exchange Program (Grant No. B08040). Q. Z. is grateful for support from the National Nuclear Security Administration under the Stewardship Science Academic Alliances program through DOE Cooperative Agreement DE-NA0001982. X. D. and X. F. Z thanks the computing resources of Tianhe II and the support of Chinese National Supercomputer Center in Guangzhou.




\begin{references}
\bibitem{R41} The initial magnetic moment was generated by random numbers (randSeed), which are classified as: (I) $0 < randSeed \leq 0.2$, it is a NM structure with the MAGMOM = 0 0 0 $\ldots$; (II) $0.2 < randSeed \leq 0.4$ it is a FM-LS structure with the MAGMOM = 1 1 1 $\ldots$; (III) $0.4 < randSeed \leq 0.6$, it is a FM-HS structure with the MAGMOM = 4 4 4 $\ldots$; (IV) $0.6 < randSeed \leq 0.8$, it is an AFM-LS structure with the MAGMOM = -1 1 -1 1 $\ldots$; (V) $0.8 < randSeed \leq 1.0$, it is an AFM-HS structure with the MAGMOM = -4 4 -4 4 $\ldots$.
\end{references}
\end{document}